\documentclass[aip,jcp,superscriptaddress,amsmath,amssymb,twocolumn,reprint]{revtex4-1}

\usepackage[version=3]{mhchem} 
\usepackage{graphicx}
\usepackage{amsmath}
\usepackage{amssymb}
\usepackage{nicefrac}
\usepackage{booktabs}
\usepackage{mwe}
\usepackage{array}
\usepackage{syntonly}

\usepackage{caption}
\usepackage{subcaption}
\usepackage{graphicx}
\newcolumntype{m}{>{$} c <{$}}
\usepackage{color}

\def\rv{{\bf r}}

\def\W{{\bf W}}
\def\X{X_{\rm corr}}
\def\dd{\mathrm{d}}

\def\beq{\begin{equation}}
\def\eeq{\end{equation}}

\def\cc{\mathrm{c}}

\def\X{\lambda_{\rm ext}}
\def\M{\rm{MAP}}

\newcommand\SV[2]{\textcolor{black}{{#2}}}
\newcommand\SVc[2]{\textcolor{black}{{#2}}}
\newcommand\SVV[2]{\textcolor{black}{{#2}}}

\begin{document}     

\author{Stefan Vuckovic}
\affiliation
{Department of Chemistry, University of California, Irvine, CA 92697, USA}
\email{svuckovi@uci.edu}
\author{Eduardo Fabiano}
\affiliation{Institute for Microelectronics and Microsystems (CNR-IMM), Via Monteroni, Campus Unisalento, 73100 Lecce, Italy}
\author{Paola Gori-Giorgi}
\affiliation{Department of Theoretical Chemistry and Amsterdam Center for Multiscale Modeling, FEW, Vrije Universiteit, De Boelelaan 1083, 1081HV Amsterdam, The Netherlands}
\author{Kieron Burke}
\affiliation
{Department of Chemistry, University of California, Irvine, CA 92697, USA}



\title{MAP: an MP2 accuracy predictor for weak interactions from adiabatic connection theory}

\begin{abstract}
Second order M\o ller-Plesset perturbation theory (MP2) approximates the exact Hartree-Fock (HF) adiabatic connection (AC) curve by a straight line.  Thus by using the deviation of the exact curve from the linear behaviour, we construct an indicator for the accuracy of MP2. We then use an interpolation along the HF AC to transform the exact form of our indicator into a highly practical MP2 accuracy predictor (MAP) that comes at negligible additional computational cost. We show that this indicator is already applicable to systems that dissociate into fragments with a non-degenerate ground state, and we illustrate its usefulness by applying it to the S22 and S66 datasets. 
\end{abstract}

\maketitle

\section{Introduction}



The adiabatic connection (AC) formalism connects a single particle picture to the fully interacting system in different electronic structure theories\cite{PauliHF,hellman1937einfuhrung,Feynman1939,HarJon-JPF-74,LanPer-SSC-75,GunLun-PRB-76,SavColPol-IJQC-03,VucWagMirGor-JCTC-15,LiuBur-PRA-09,VucLevGor-JCP-17,Per-IJQC-18,Per-PRL-18}. As such, it has played an important role in the development of both density functional theory (DFT) and wavefunction theory (WFT) methods. On the DFT side, the AC provides justification and rationalization of widely popular hybrid\cite{Bec-JCP-93,PerErnBur-JCP-96,ZhaSchTru-JCTC-06}  and double hybrid functionals, \cite{Gri-JCP-06,LarGri-JCTC-10,ShaTouSav-JCP-11}  and it has been used for the construction of other classes of density functional approximations.\cite{Ern-CPL-96,SeiPerKur-PRL-00,MorCohYan-JCP-06,Bec-JCP-13a,VucIroSavTeaGor-JCTC-16,VucIroWagTeaGor-PCCP-17,BahZhoErn-JCP-16,VucGor-JPCL-17,GouVuc-JCP-19,Vuc-JCTC-19} A simple geometric construction of the AC curve has been used to obtain a lower bound to the correlation energy in DFT,\cite{VucIroWagTeaGor-PCCP-17} and it has been used to rationalize the amount of exact exchange in the widely used PBE0 hybrid functional.\cite{BurErnPer-CPL-97,PerErnBur-JCP-96} 
On the WFT side, the Hartree-Fock (HF) AC has as weak-interaction expansion the M\o ller-Plesset perturbation theory\cite{MolPle-PR-34}. It was also recently proposed how the AC formalism can be used to recover missing correlation energy for a broad range of multireference WFTs.\cite{Per-PRL-18,PasPer-JCTC-18,Per-JCP-18}

In the present paper, we use the AC formalism to gain more insight into the performance of second-order perturbation theory and provide an indicator for its accuracy. Our construction is very simple and uses the fact that in the second-order perturbation theories (PT2; in both DFT and HF variants of the AC formalism) the AC curve is approximated by a straight line, whose slope is equal to twice the PT2 correlation energy. Thus,  the two PT2 are more accurate the more linear the exact AC curve is. Following this, a remarkably simple geometric construction of the AC curves yields an indicator for the accuracy of the PT2 methods. 
We use an interpolation along the HF adiabatic connection formalism to transform the exact form of our indicator into a practical tool for predicting the accuracy of MP2. We show that this tool is readily applicable to systems that dissociate into fragments with nondegenarate ground states. Applying it to the S22 and S66 datasets, we illustrate the usefulness of our indicator for predicting failures of MP2 when applied to noncovalently bonded systems.  

\section{Theory}

We briefly review the basics of the AC formalism in DFT and HF theory. In either theory, we define a coupling-constant $\lambda$ dependent Hamiltonian.  In DFT, it reads as:\cite{HarJon-JPF-74,LanPer-SSC-75,GunLun-PRB-76}
\beq\label{eq:adiabDFT}
	\hat{H}_{\lambda}^{\rm DFT}=\hat{T}+\lambda\,\hat{V}_{ee}+\hat{V}_{\lambda}^{\rm DFT},
\eeq
where $\hat{T}$ is the kinetic energy operator and $\hat{V}_{ee}$ is the electron-electron repulsion operator. The $\hat{V}_{\lambda}^{\rm DFT}$ operator represents a one-body potential, which forces $\Psi_\lambda^{\rm DFT}$, the ground state of eq~\ref{eq:adiabDFT}, to integrate to the physical density $\rho=\rho_1$ for all $\lambda$ values. At $\lambda=1$,  $\hat{V}_{\lambda}^{\rm DFT}$ is equal to $\hat{V}_{\rm ext}$ the (nuclear) external potential. The corresponding HF AC Hamiltonian is given by (see, e.g., refs~\onlinecite{Per-IJQC-18} and ~\onlinecite{SeiGiaVucFabGor-JCP-18}): 
\begin{equation}\label{eq:HlambdaHF}
	\hat{H}_{\lambda}^{\rm HF}=\hat{T}+\hat{V}_{\rm ext}+\lambda \hat{V}_{ee}+
	\Big(1-\lambda \Big)
	\Big(\hat{J}+\hat{K} \Big),
\end{equation}
where $\hat{J}=\hat{J}[\rho^{\rm HF}]$ and $\hat{K}=\hat{K}[\{\phi_i^{\rm HF}\}]$ are the standard HF Coulomb and exchange operators that depend on the HF density $\rho^{\rm HF}$ and occupied HF orbitals $\phi_i^{\rm HF}$. They are computed once in the HF calculation for the physical system and do not depend on $\lambda$. 
A key difference between the two ACs is that the density of $\Psi_{\lambda}^{\rm HF}$ (the ground state of the Hamiltonian of eq~\ref{eq:HlambdaHF}) varies with $\lambda$, whereas the density of $\Psi_{\lambda}^{\rm DFT}$  is always forced to be that of the physical system. But at $\lambda=1$, $\hat{H}_{1}^{\rm DFT}= \hat{H}_{1}^{\rm HF}= \hat{H}$, and thus: $\Psi_{1}^{\rm DFT}= \Psi_{1}^{\rm HF}=\Psi$. 
\begin{figure*}
\centering
 \begin{subfigure}[t]{0.45\textwidth}
 \centering
  \includegraphics[width=8cm]{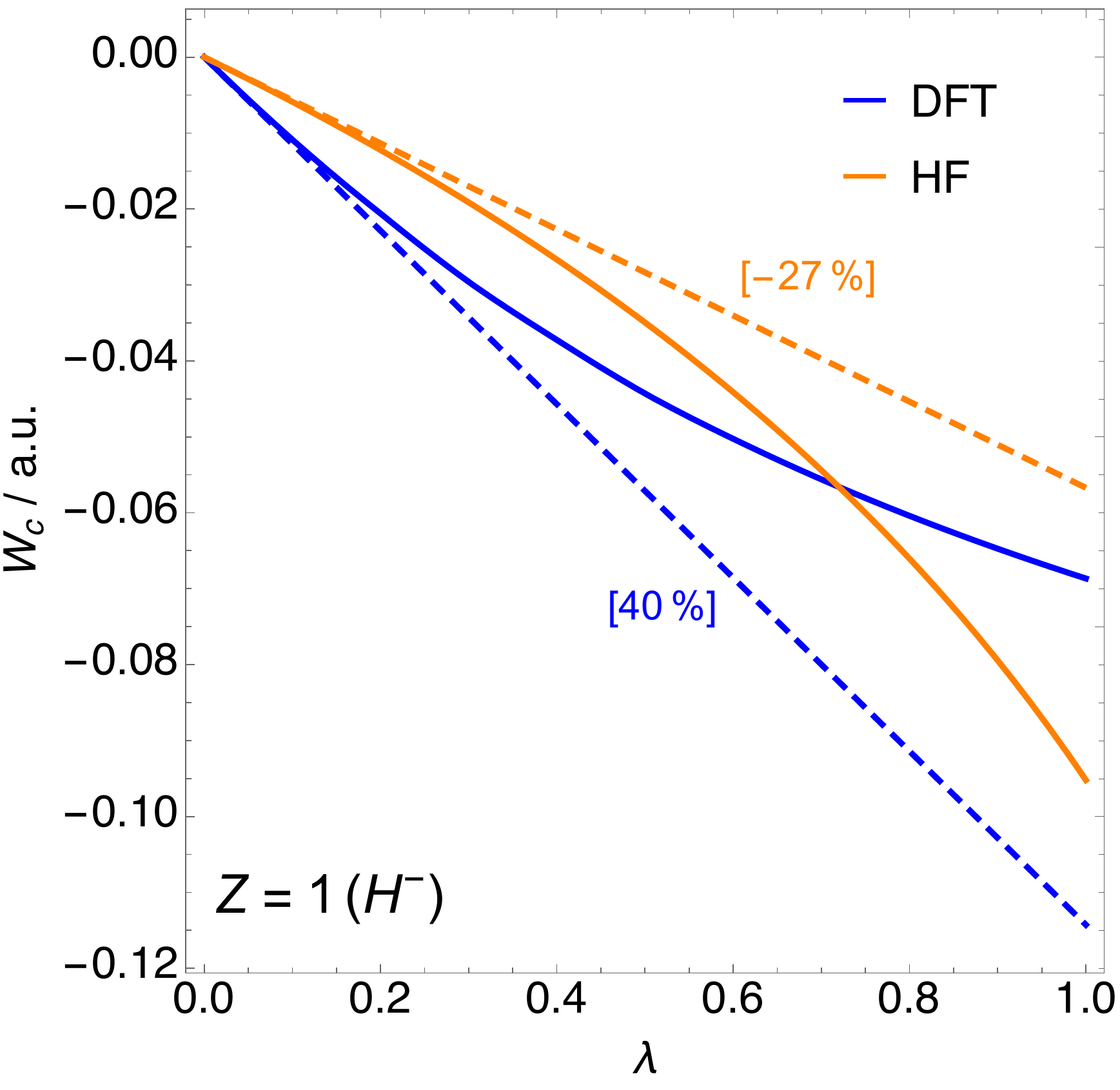}\par
   \includegraphics[width=8cm]{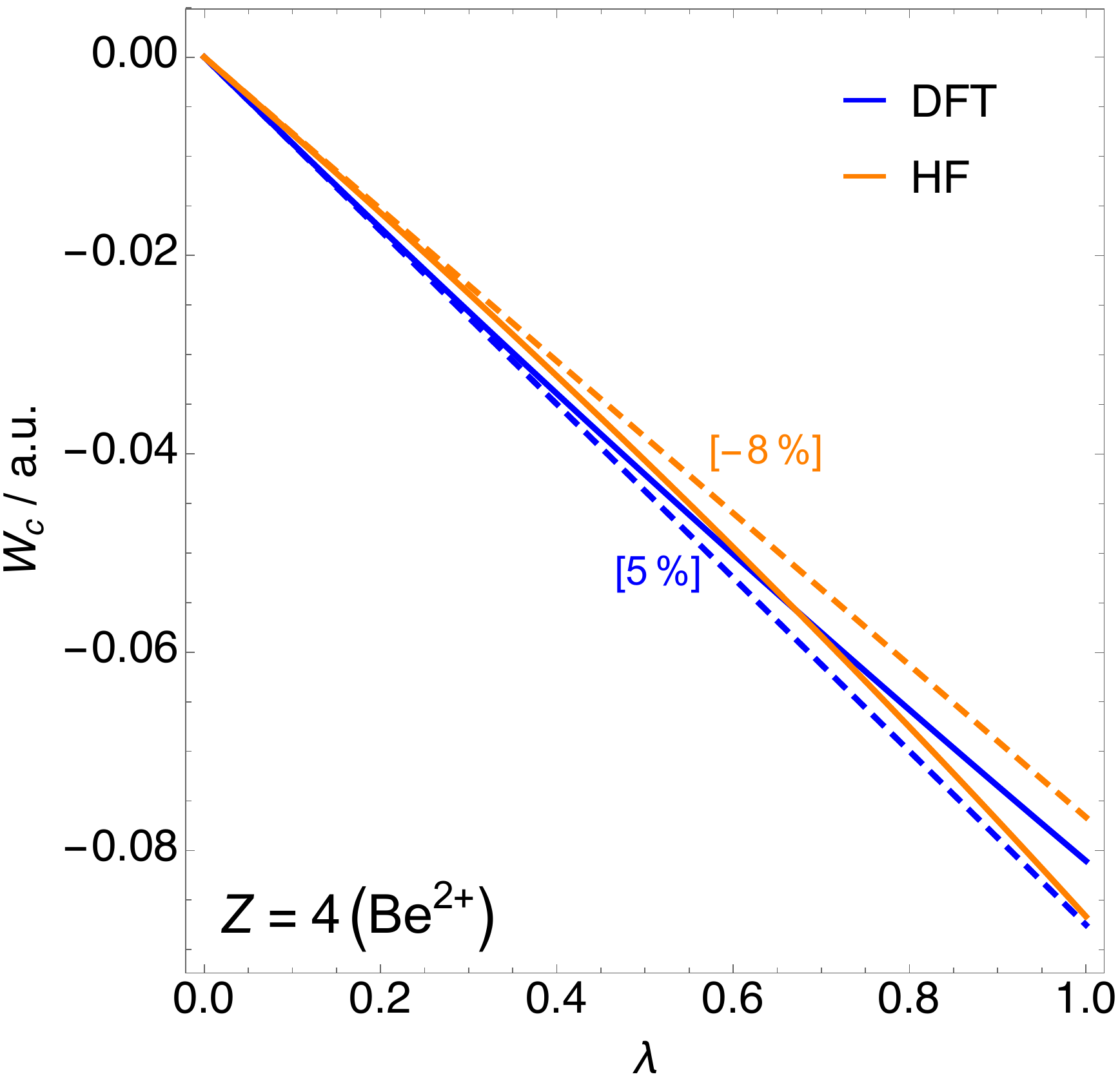}\par
 \end{subfigure}
 \begin{subfigure}[t]{0.45\textwidth}
 \centering
  \includegraphics[width=8cm]{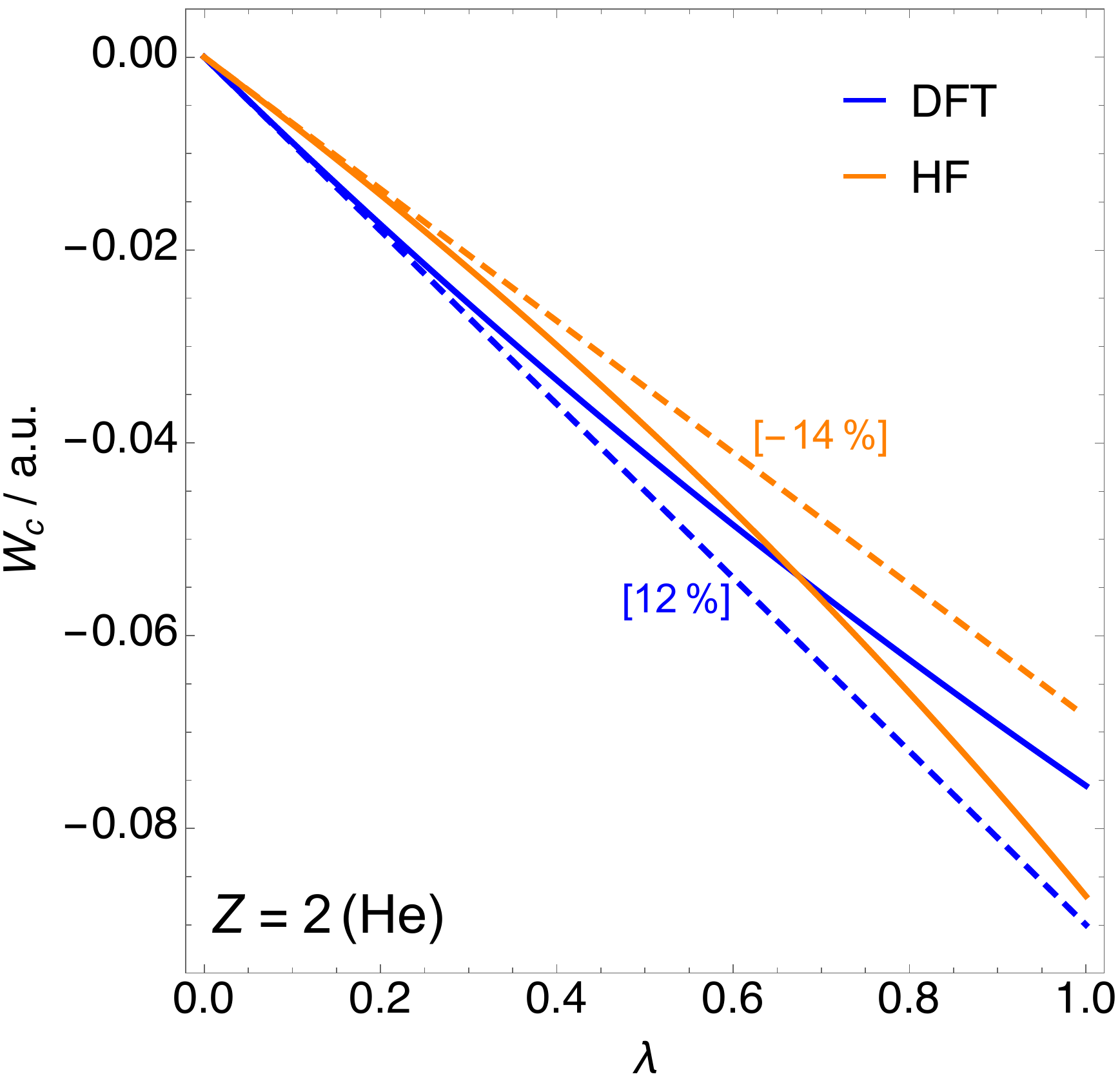}\par
    \includegraphics[width=8cm]{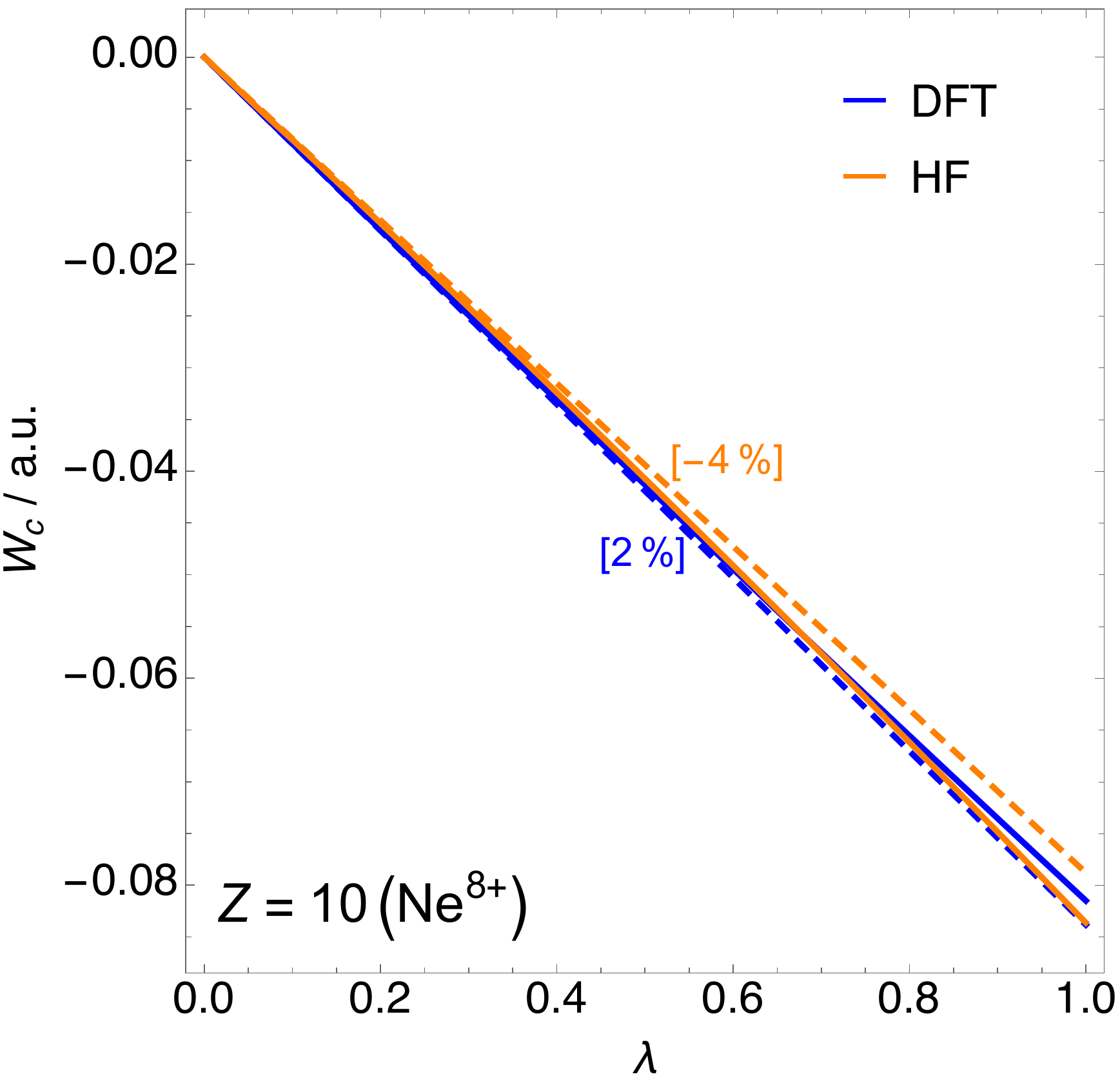}\par
 \end{subfigure}
\caption{DFT and HF AC curves for the selected members of the helium isoelectronic series. Dashed lines represent the AC curves from the second-order perturbation theory $2 E_c^{{\rm PT}2}\,\lambda$. \SV{}{The numbers in square brackets are the relative errors of $E_c^{{\rm PT}2}$:
$\left(E_c^{\rm MP2}-E_{\rm c}^{\rm HF} \right) / E_{\rm c}^{\rm HF}$ in the case of HF AC, and $\left(E_c^{\rm GL2}-E_{\rm c}^{\rm DFT} \right) / E_{\rm c}^{\rm DFT}$ in the case of DFT AC }
}
\label{fig_hhe}
\end{figure*}

\begin{figure}
 \includegraphics[width=0.95\linewidth]{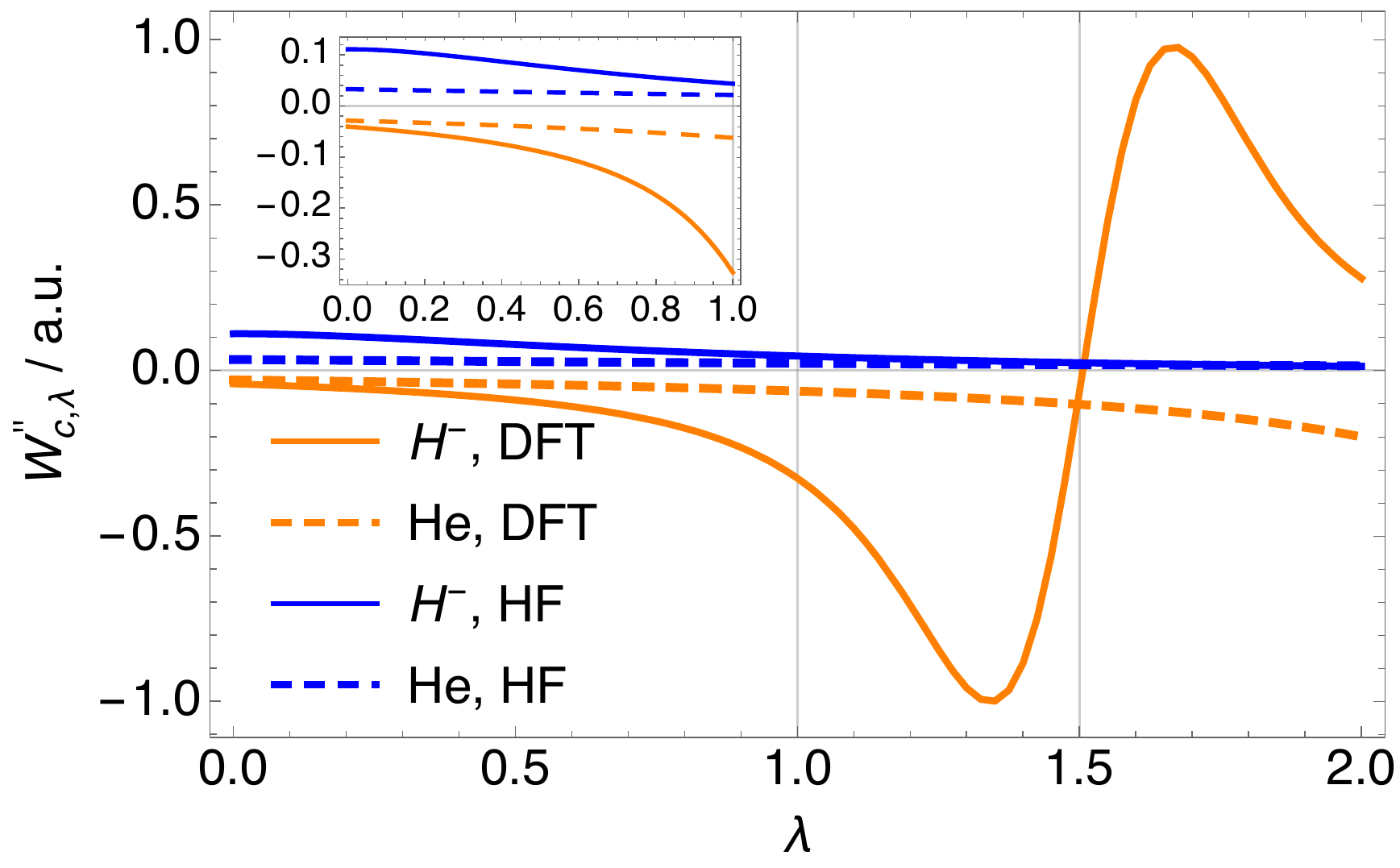}\par
\caption{The curvature of the HF and DFT AC curves, $W''_{c,\lambda}= \partial^2 W_{c,\lambda}/ \partial \lambda^2$, 
 for H$^-$ and He.\SV{}{The inset zooms in on the region of the plots for  the $\lambda$ domain between $0$ and $1$. The DFT AC fitting functions used in this figure are given in supporting information}} 
\label{fig_second}
\end{figure}
In either theory,
\beq
E_{\rm c}=\langle  \Psi| \hat{H} | \Psi  \rangle - \langle  \Psi_0 | \hat{H} | \Psi_0 \rangle, 
\eeq
and the AC formula for the correlation energy follows in both cases from the Hellmann–Feynman theorem,
\beq \label{eq:ac}
E_{\rm c}= \int_0^1 W_{\cc,\lambda} \dd \lambda.
\eeq
In DFT, the underlying AC integrand $W_{\cc,\lambda}$ is  given by
\beq \label{eq:wcdft}
W_{\cc, \lambda}=\langle \Psi_\lambda
|\hat{V}_{\rm ee} |
\Psi_\lambda \rangle 
- \langle  \Psi_0
 |\hat{V}_{\rm ee} 
 |  \Psi_0 \rangle~~~\text{(DFT)},
\eeq
whereas its HF counterpart is 
\beq \label{eq:wchf}
W_{\cc, \lambda}=\langle \Psi_\lambda
|\hat{V}_{\rm ee} - \hat{J}-\hat{K}|
\Psi_\lambda \rangle 
- \langle  \Psi_0
 |\hat{V}_{\rm ee} - \hat{J}-\hat{K}
 |  \Psi_0\rangle ~~~\text{(HF)}.
\eeq
In DFT (eq~\ref{eq:wcdft}), $\Psi_0$ is the Kohn-Sham wavefunction, and in the HF AC (eq~\ref{eq:wchf}), $\Psi_0$ is the HF Slater determinant, which minimizes $\hat {H}$. 
Utilizing the expansion of $W_{\lambda,c}^{\rm DFT/HF}$ at small $\lambda$ \SVV{}{up to $n$-th order }, we can write:
\beq
W_{\lambda,c}^{(n)} = \sum_{m=2}^{n} m\,E_c^{{\rm PT}m}\,\lambda^{m-1} 
,
 \label{eq:lambda0DFT} 
\eeq
where $E_c^{{\rm PT}n}$ is the correlation energy from the $n$-th order perturbation theory, given by
\beq
E_{\rm c}^{(n)}=  \sum_{m=2}^{n} E_c^{{\rm PT}m} 
.
\eeq
 Within the HF AC, $E_c^{{\rm PT}m}$  is obtained from M\o ller-Plesset (MP) perturbation theory (PT$=$MP), whereas in the DFT case $E_c^{{\rm PT}m}$  is obtained from G\"orling-Levy perturbation theory (PT$=$GL).\cite{GorLev-PRB-93,GorLev-PRA-94} By truncation to second order in $\lambda$, $W_{\lambda,c}$ is approximated by a straight line: 
\beq \label{eq:lin}
W_{\lambda,c}^{\rm DFT/HF} \approx 2 E_c^{{\rm PT}2}\,\lambda,
\eeq
which sets $E_{\rm c}^{\rm DFT / HF} \approx E_c^{{\rm PT}2} $. Both MP2 and GL2 theories are pillars of electronic structure theory, and their use is widespread in many calculations. 
Besides the widespread use of the MP2 method and its extensions in their standalone versions (see, e.g., ref~\onlinecite{Cre-WIR-11} for a review), the PT2 correlation energy is also used as an ingredient for double hybrids\cite{Gri-JCP-03,JunLocDutHea-JCP-04,NeeSchKosSchGri-JCTC-09}.

\section{Illustrations}



In Figure~\ref{fig_hhe}, we show the AC curves in DFT and HF theories for the members of the helium isoelectronic series, namely for H$^-$, He, Be$^{2+}$ and Ne$^{8+}$. 
For H$^-$ and He, the second derivative of both $W_{c,\lambda}^{\rm HF}$ and $W_{c,\lambda}^{\rm DFT}$ (w.r.t.~$\lambda$) is plotted in Figure~\ref{fig_second} for $\lambda$ values between $0$ and $2$. 
The AC curves have been obtained from the $\Psi_{\lambda}^{\rm HF/DFT}$ wavefunctions at the full-CI/aug-cc-pCVTZ level.\cite{Dun-JCP-89} The DFT AC curves have been taken from Refs.~\onlinecite{locpaper,TeaCorHel-JCP-09}, while those of the HF AC have been obtained from the $\Psi_\lambda^{\rm HF}$ wavefunction, which we construct in the present work [the full details are given in the supporting information]. 
While both AC curves decrease with $\lambda$, that their convexity can be different is already evident from Figure~\ref{fig_hhe}. \SV{}{As it can be seen from Figure~\ref{fig_second},   $W_{\lambda,c}^{\rm DFT}$ is convex for both systems.} In fact,  $W_{\lambda,c}^{\rm DFT}$ is believed to be always convex (or at least piecewise convex)\cite{VucIroWagTeaGor-PCCP-17} and this is supported by the highly accurate numerical evidence.\cite{TeaCorHel-JCP-09,TeaCorHel-JCP-10,locpaper} On the other hand, we can see from Figure~\ref{fig_second} that the convexity of $W_{\lambda,c}^{\rm HF}$ is not definite.   For H$^-$, $W_{\lambda,c}^{\rm HF}$ is concave up to $\lambda \sim 1.5$ and then it becomes convex. For He, the convexity changes later, at $\lambda \sim 3.4$.  
In fact, although often concave at small $\lambda$, we know that $W_{\lambda,c}^{\rm HF}$ must change convexity at larger $\lambda$, in order to approach a finite asymptotic value\cite{SeiGiaVucFabGor-JCP-18} $W_{\infty,c}^{\rm HF}$ when $\lambda \to \infty$. 

Staying with Figures~\ref{fig_hhe}, we can notice that the curvature of both DFT and HF AC curves are the strongest in the case of H$^-$, and then it decreases as we increase the nuclear charge, $Z$. Thus the relative errors in the corresponding GL2/MP2 correlation energies also decrease with $Z$ (even though the GL2 overestimates here the magnitude of $E_c^{\rm DFT}$ and MP2 underestimates the magnitude of $E_c^{\rm HF}$ in all cases). Furthermore, the DFT and HF curves are getting closer to each other as $Z$ increases, and for Ne$^{8+}$ the two curves are nearly overlapping.

\section{Practical predictor for the accuracy of the MP2 theory when applied to noncovalent systems}\label{sec:predictor}

\begin{figure}
 \includegraphics[width=0.95\linewidth]{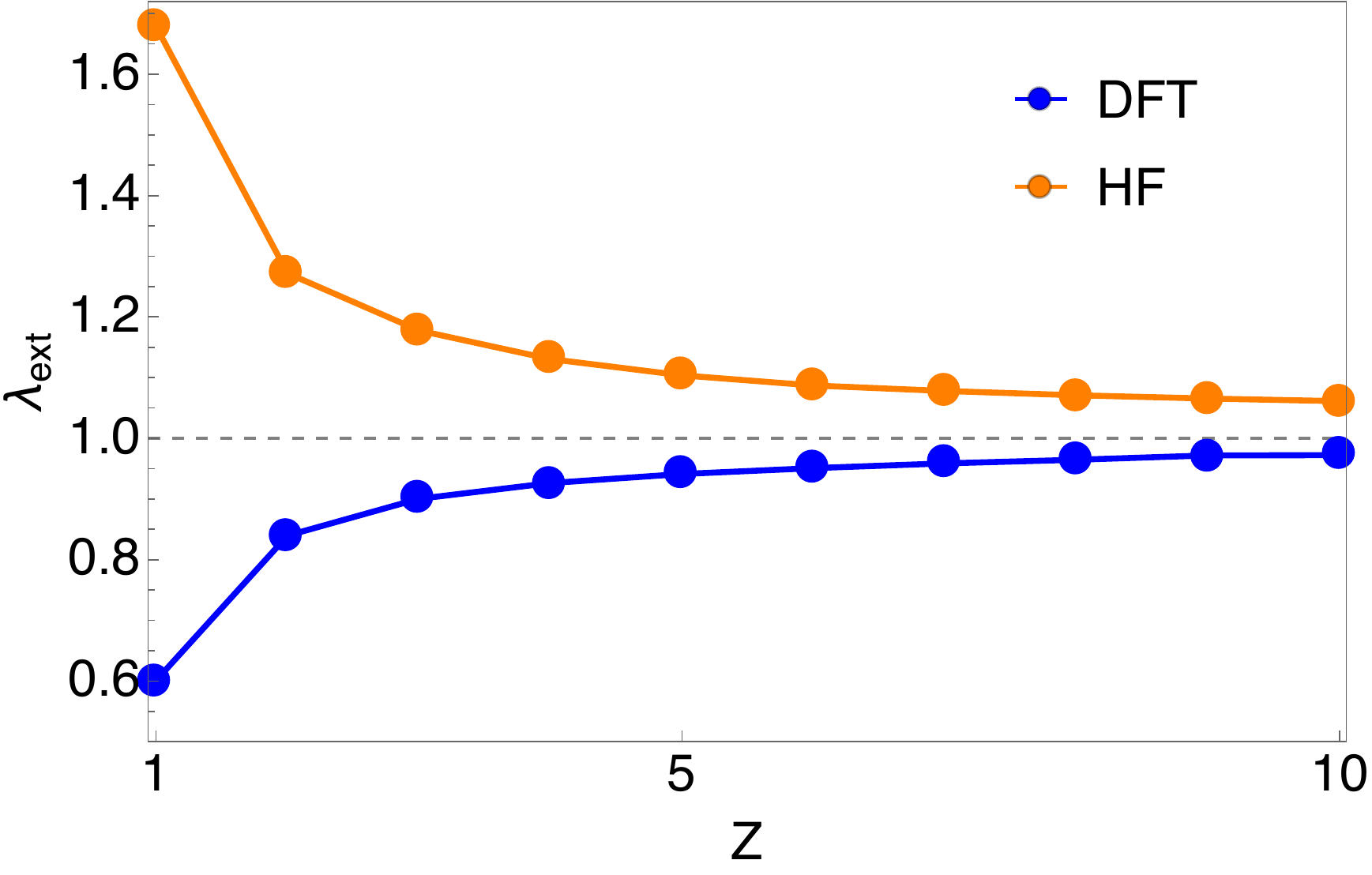}\par
\caption{$\X$ for both DFT and HF AC curves for the helium isoelectronic series as a function of nuclear charge, $Z$} 
\label{fig_Xhe}
\end{figure}

Utilizing that
$E_c^{\rm PT2}$ is more accurate the more linear the exact $W_{c,\lambda}$ is, here we use a quantity defined in ref~\onlinecite{VucIroWagTeaGor-PCCP-17} as an indicator of accuracy of the two second-order perturbation theories. This indicator is defined by:\cite{VucIroWagTeaGor-PCCP-17} 
\beq
\label{eq:X}
\X =\frac{W_{c,1}}{W'_{c,0} }. 
\eeq
The $\X$ quantity is simply
a value of $\lambda$ at which the extrapolated PT line ($W_{c,\lambda}=\lambda W'_0=2\lambda E_{c}^{\rm PT2} $ )  reaches 
the $W_{c,\lambda}=W_{c,1}$ horizontal line.
As such, it represents a dimensionless measure of the curvature of ACs. For ACs convex in $\lambda$ (within the relevant $\lambda$ region between $0$ and $1$), $\X $ needs to be less than $1$. For these curves, the error of PT2 vanishes as $\X$ approaches $1$ (from below). Thus, when the AC curve is a straight line, $X$ is equal to $1$ and the PT2 is exact. For AC curves concave in $\lambda$ (again within the relevant $\lambda$ region between $0$ and $1$), $\X $ is greater than $1$ and for these AC curves the error of PT2 also vanishes when $\X$ approaches $1$ (from above). To illustrate this, in Figure~\ref{fig_Xhe}, we show $\X$ for the members of the helium isoelectronic series. In the case of HF ACs, the underlying $\X$ value for H$^-$ is $\sim 1.7$, for He it drops to $\sim 1.3$ and then it further decreases with $Z$.
In the case of DFT ACs, the underlying $\X$ value for H$^-$ is $\sim 0.6$, then for He it increases  to $\sim 0.8$ and then further increases with $Z$. 
In both DFT and HF case, $\X$ approaches $1$ (although from different directions)  as $Z$ increases, 
given that both MP2 and GL2 correlation energies become exact in the $Z$ limit of the helium isoelectronic series.\cite{GorLev-PRA-94,locpaper} \SV{}{We can also see from Figure~\ref{fig_Xhe} 
that $\X$  pertaining to the DFT AC approaches $1$ faster than its HF counterpart. This mirrors the fact the error of GL2 error decreases more quickly than that of MP2 at larger $Z$ (Figure~\ref{fig_hhe}).
}

So far we have discussed the differences between the HF and DFT AC curves and in the remainder of this paper we focus only on HF AC aiming to provide a practical tool for predicting the accuracy of MP2.
The quantity of eq~\ref{eq:X}, via $W_{c,1}$, requires knowledge of the fully interacting wave-function, and 
thus its direct use as an indicator for the accuracy of MP2 is impractical. We aim at circumventing this problem by obtaining $W_{c,1}^{\rm HF}$ via interpolation between the weakly and strongly interacting limits of the ACs. This idea was proposed by Seidl and co-workers in the context of the DFT AC.\cite{SeiPerLev-PRA-99,SeiPerKur-PRL-00} Recent papers have also explored its use in the context of the HF AC, obtaining rather good results for interaction energies, particularly\cite{FabGorSeiSal-JCTC-16,VucGorDelFab-JPCL-18} (but not only\cite{GiaGorDelFab-JCP-18}) of non-covalently bonded systems.  To use this approach in the HF AC context, we employ the following SPL (after Seidl, Perdew and Levy) interpolation form\cite{SeiPerLev-PRA-99}
\beq
\label{eq:SPL}
 W_{c,\lambda}^{\rm SPL}(\mathbf{W})=
W_{c,\infty} \left( 1- \left( 1 + \frac{4  E_c^{\rm PT2}\lambda }{W_{c,\infty} } \right)^{-1/2} \right),
\eeq
where $\mathbf{W}=\{W^1,...,W^k\}$ is the set of input ingredients from which the interpolation is built, which in this case is $\mathbf{W}=\{W_0,W'_0,W_\infty\}$, with $W_0=E_x$, $W_0'=2E_c^{\rm PT2}$, $W_{c,\infty}=W_\infty-W_0$. This interpolation form has been used extensively in the literature.\cite{SeiGorSav-PRA-07,VucIroWagTeaGor-PCCP-17,VucIroSavTeaGor-JCTC-16,KooGor-TCA-18,Vuc-JCTC-19} We should immediately remark that $W_{c,\lambda}^{\rm SPL}$ is always convex, and as such cannot provide a good model for the HF adiabatic connection of a given system. However, as most often in chemistry, we are interested here in interaction energies. At least for non-covalently bonded systems, interpolations like the SPL one for interaction energies work extremely well in the HF case,\cite{FabGorSeiSal-JCTC-16,VucGorDelFab-JPCL-18} pointing to the fact that the interaction energy HF adiabatic connection curve is probably convex, and very well modeled by the difference between two convex curves, as we are going to detail in the following. 


\begin{figure}
    \includegraphics[width=0.95\linewidth]{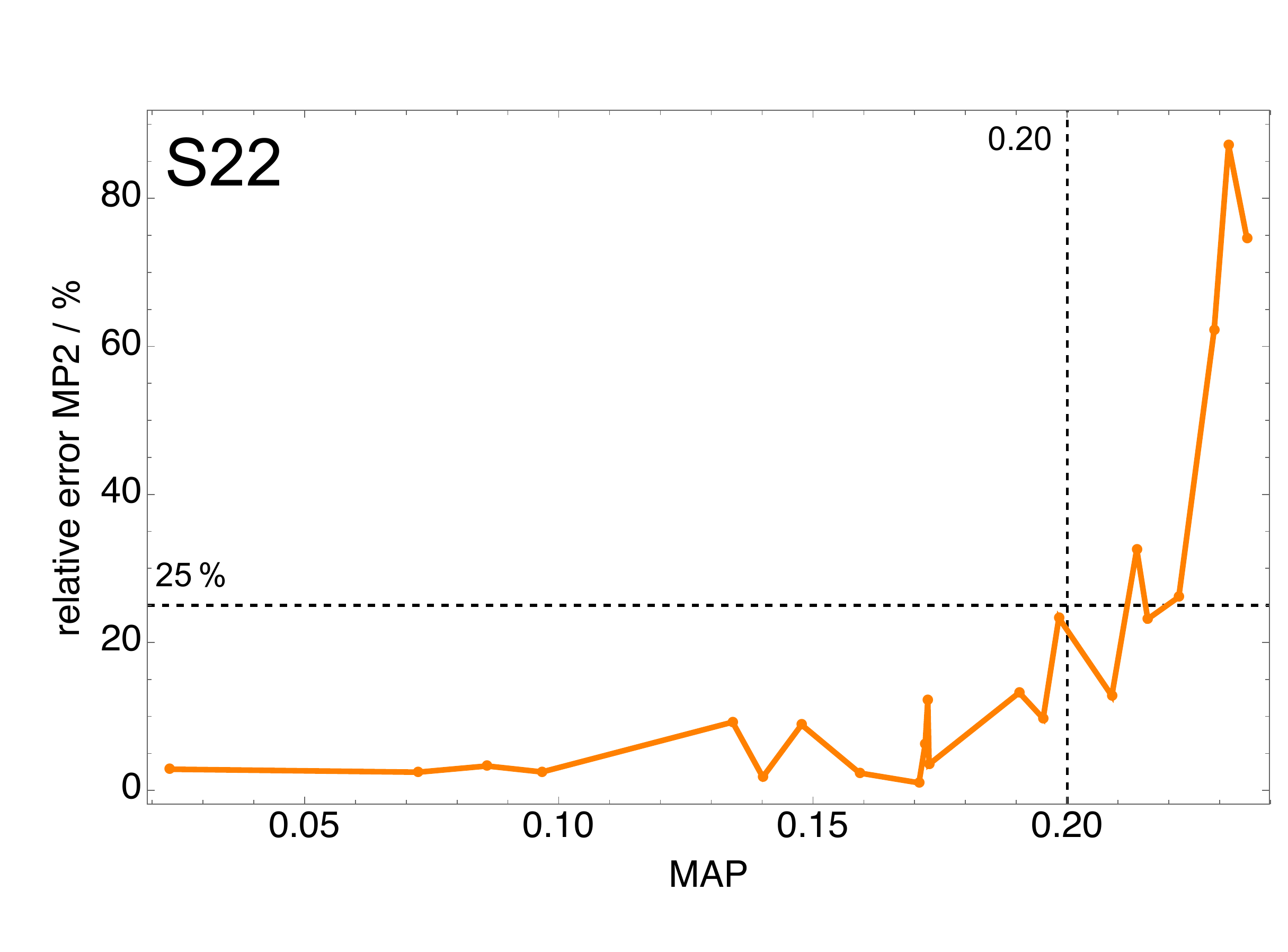}\par
    \includegraphics[width=0.95\linewidth]{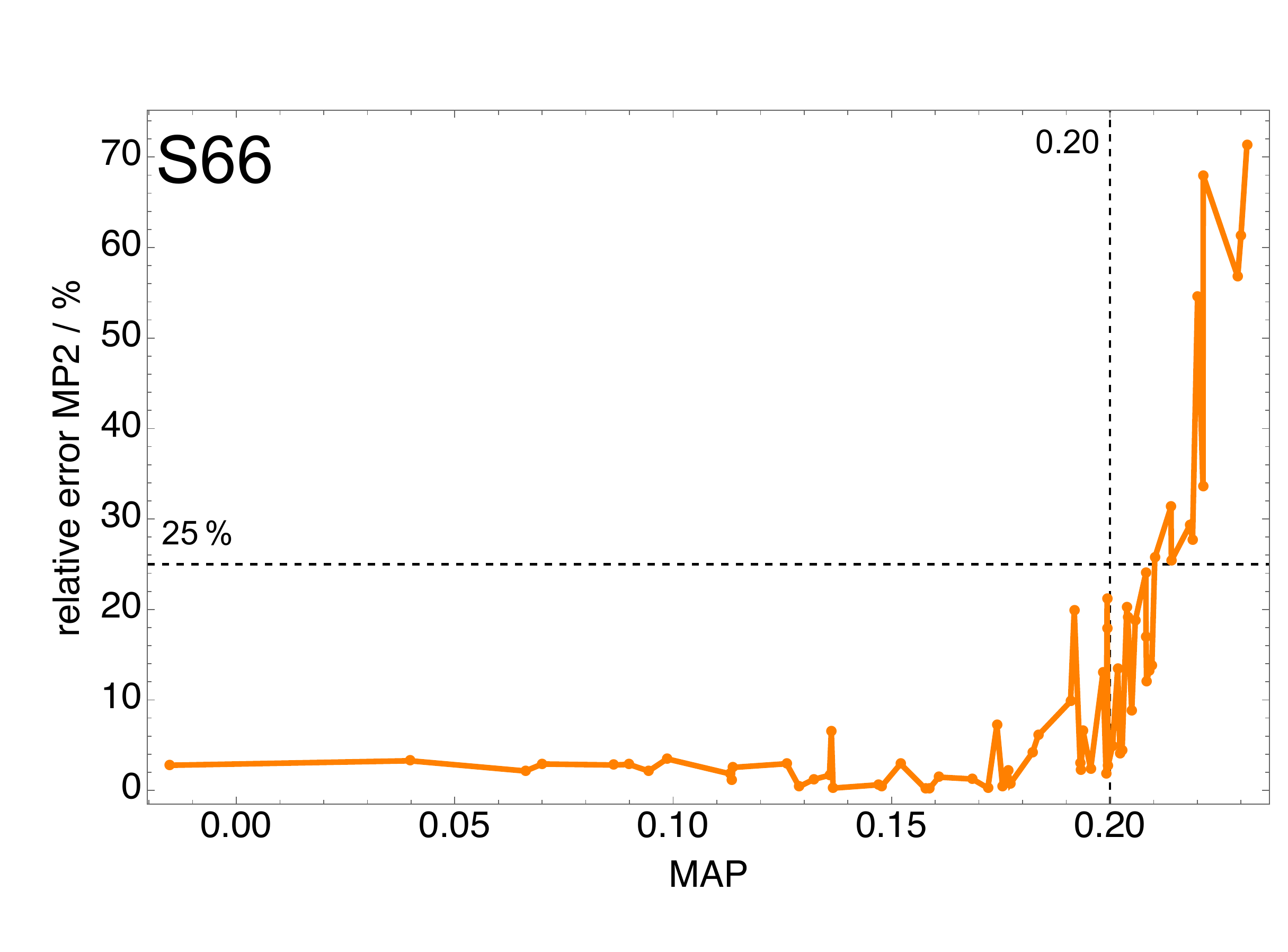}\par
\caption{The relative errors in MP2 binding energies as a function of $\X^{\rm SPL}$ for the S22 (top panel) and S66 (lower panel) datasets}
\label{fig_S}
\end{figure}

Consider a bound system (e.g., a molecular complex) $M$ whose individual fragments are $F_i$. We are interested in the interaction energy AC curve, which is given by: 
\beq \label{eq:e}
W_{\lambda,c}^{\rm int}(M)=W_{\lambda,c}(M) - \sum_{i=1}^{N} W_{\lambda,c} (F_i).
\eeq
To compute $ W_\lambda^{\rm SPL,int}(M)$, we generalize the size-consistency correction of ref~\onlinecite{VucGorDelFab-JPCL-18} to define:
\beq \label{eq:wcspl}
W_{c,\lambda}^{\rm SPL,int}(M)= W_{c,\lambda} ^{\rm SPL} \left(    \W(M)         \right)     
 -
 W_{c,\lambda} ^{\rm SPL} \left( \sum_{i}^N   \W(F_i)         \right),
\eeq
where $ \W(M) $ and $\W(F_i)$ are the input ingredients of the complex and of the fragments, respectively.
Equation~\ref{eq:wcspl} works for a system $M$ whose fragments $F_i$ have nondegenerate ground states, as in this case it is guaranteed that $W_{c,\lambda}^{\rm SPL,int}(M)$ vanishes when the distance between the fragments is set to infinity. The use of eq~\ref{eq:wcspl} is discussed in more details in supporting information.
 \begin{figure}
 \includegraphics[width=0.95\linewidth]{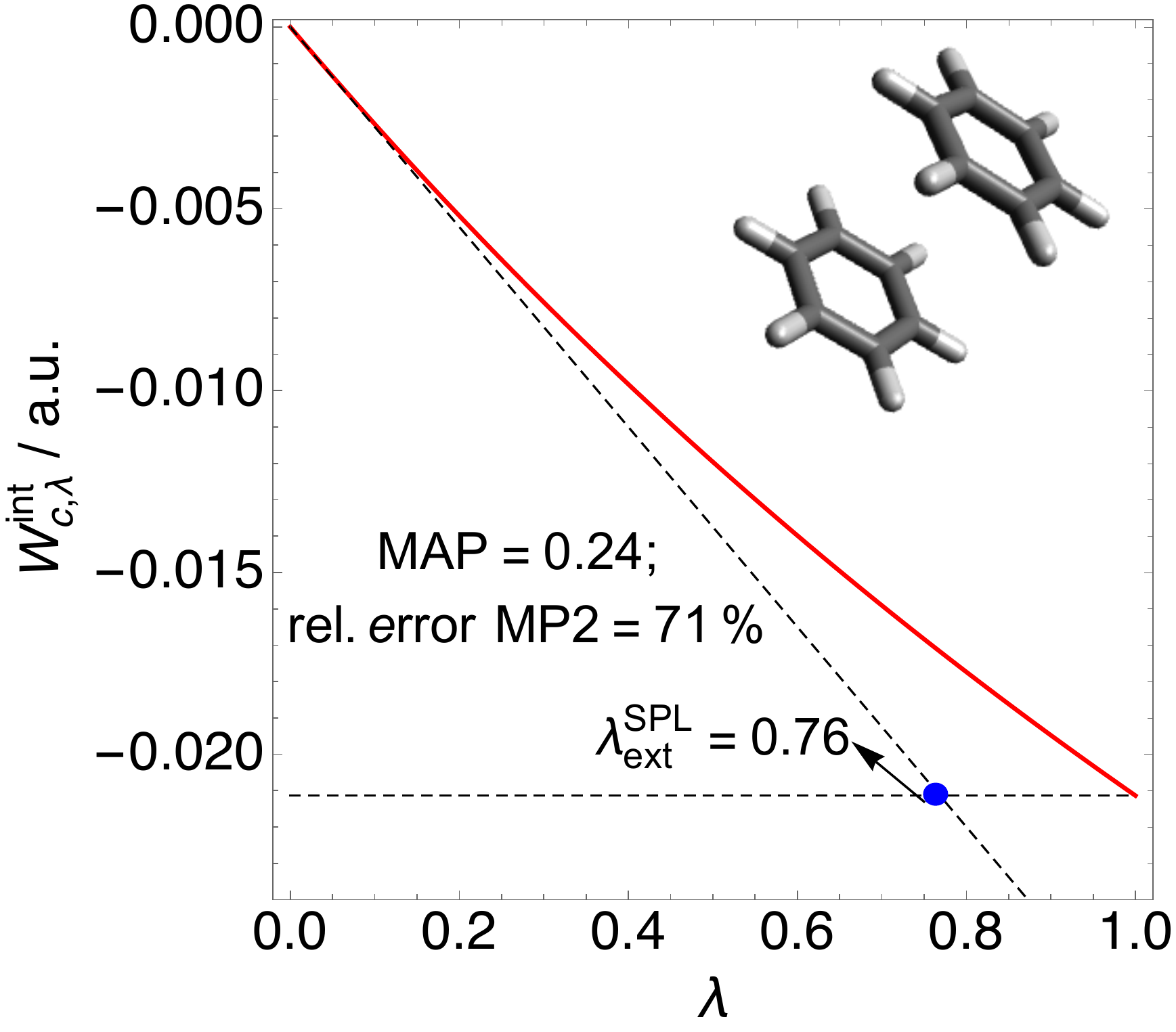}\par
    \includegraphics[width=0.95\linewidth]{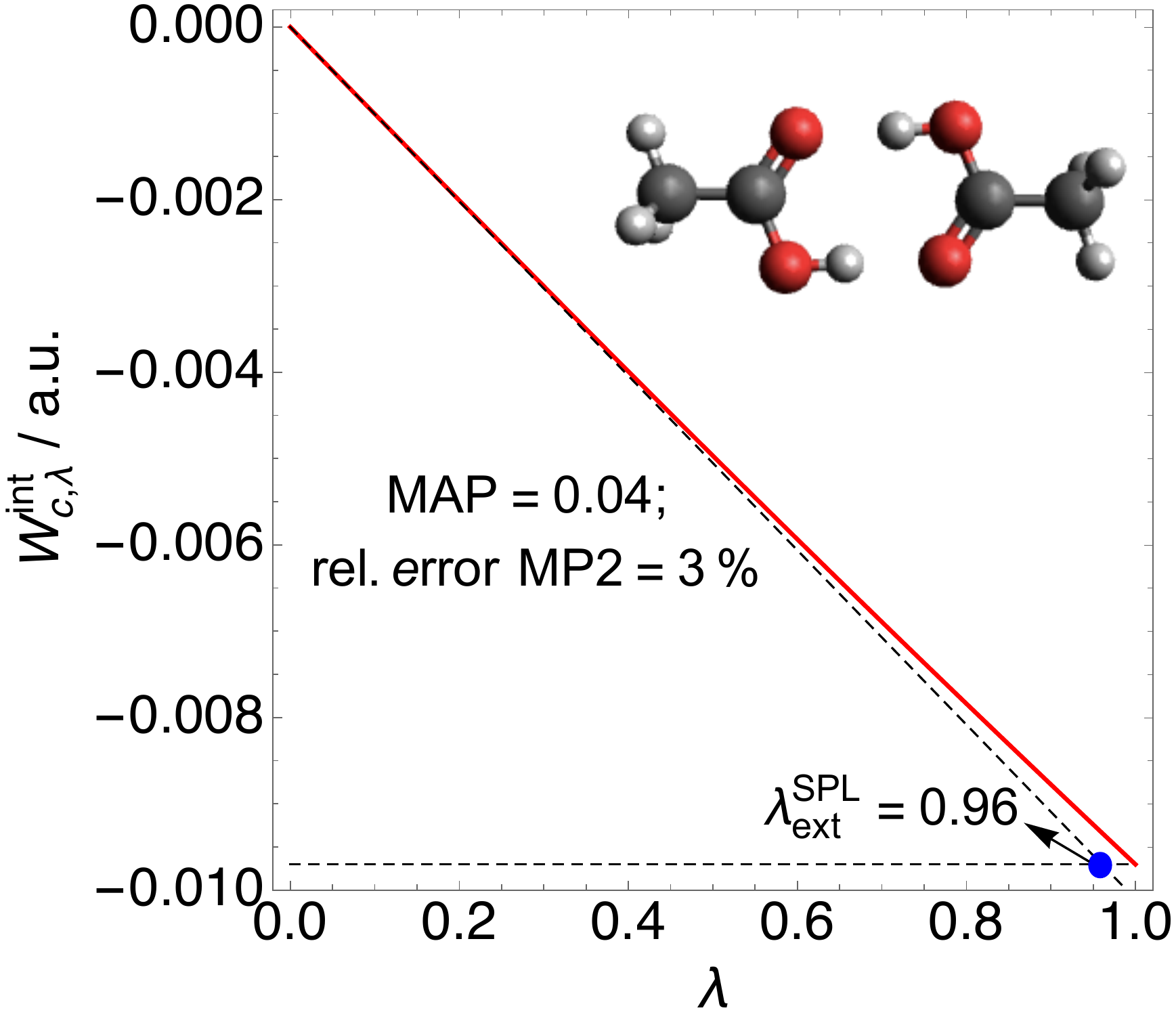}\par
\caption{The interaction AC curves obtained by the SPL interpolation via eq~\ref{eq:SPL} for the benzene dimer (upper panel) and the acetic acid dimer (lower panel)} 
\label{fig_C6}
\end{figure}

To complete the model, we need $W_\infty^{\rm HF}$, whose exact (fully nonlocal) form 
has been recently revealed,\cite{SeiGiaVucFabGor-JCP-18} with ongoing  efforts in exploring whether this form can be actually useful for building approximations to $E_{c}^{\rm HF}$.  For practical reasons here we approximate $W_{\infty}^{\rm HF}$  with the point-charge-plus-continuum
(PC) semilocal model evaluated on $\rho^{\rm HF}(\rv)$:\cite{SeiPerKur-PRA-00}\SVc{}{
\beq \label{eq:pc}
W_{\infty}^{\rm HF} \approx W_{\infty}^{\rm PC}[\rho^{\rm HF}]  = \int \left[A\rho^{\rm HF}(\mathbf{r})^{4/3} + B\frac{|\nabla \rho^{\rm HF}(\mathbf{r})|^2}{\rho^{\rm HF}(\mathbf{r})^{4/3}}\right]\mathrm{d}\mathbf{r},
\eeq
where $A=-1.451$, $B=5.317\times10^{-3}$. The correlation part of $W_{\infty}^{\rm PC}$ is obtained as $W_{c,\infty}^{\rm PC}[\rho^{\rm HF}] = W_{\infty}^{\rm HF}[\rho^{\rm HF}] - E_{\rm x}^{\rm HF}$.}
It has been recently shown that the combination of the PC model approximation and the SPL interpolation form of eq~\ref{eq:wcspl} yields rather accurate interaction energies for systems that we consider in the present work.\cite{VucGorDelFab-JPCL-18} In Appendix~\ref{sec_pc}, we further discuss the use of the PC model in this context. We remark that in addition to the SPL form, other forms have been proposed in the literature (see, e.g., refs~\onlinecite{VucGorDelFab-JPCL-18} and~\onlinecite{VucIroSavTeaGor-JCTC-16}).  However, for the systems that we consider here, the difference between the results obtained with the SPL form and other ones is very small.\cite{VucGorDelFab-JPCL-18} 

\SVc{}{Combining eqs~\ref{eq:X},~\ref{eq:SPL} and~\ref{eq:wcspl} we find the $\X^{\rm SPL}$ indicator that pertains to the interaction HF AC curve of eq~\ref{eq:wcspl}:
\beq  \label{eq:x1spl}
\X^{\rm SPL}=\frac{W_{c,\lambda=1}^{\rm SPL,int}(M)}{2E_c^{\rm MP2}(M)-2 \sum_{i}^N E_c^{\rm MP2}(F_i)},
\eeq
where $W_{c,\lambda=1}^{\rm SPL,int}(M)$ is given by:
\begin{align}
\label{eq:1spl}
&W_{c,\lambda=1}^{\rm SPL,int}(M) =
W_{c,\infty}^{\rm PC}(M) \left( 1- \left( 1 + \frac{4  E_c^{\rm MP2}(M) }{W_{c,\infty}^{\rm PC}(M) } \right)^{-1/2} \right) \nonumber  \\ 
&- \sum_{i}^N W_{c,\infty}^{\rm PC}(F_i) \left( 1- \left( 1 + \frac{4    \sum_{i}^N E_c^{\rm MP2}(F_i) }{ \sum_{i}^N W_{c,\infty}^{\rm PC}(F_i) } \right)^{-1/2} \right) .
\end{align} }

\SVc{}{With eqs~\ref{eq:x1spl} and \ref{eq:1spl} we have what we need to compute $\X^{\rm SPL}$ corresponding to the HF AC for the interaction energies of molecular complexes bonded by non-covalent interactions.} The principal point of $\X^{\rm SPL}$ is its use as an indicator for the accuracy of the MP2 theory.  \SV{}{We define the MP2 accuracy predictor ($\M$) in terms of $\X^{\rm SPL}$ of eq~\ref{eq:x1spl}: $\M=1-\X^{\rm SPL}$,  to make the MP2 error increase as the predictor increases.} In Figure~\ref{fig_S}, we plot the relative error in the MP2 binding energies as a function of $\M$ for the S22 and S66 datasets.

We can observe a general trend that the MP2 errors on average decrease 
as $\M$ approaches $0$ (i.e. the corresponding AC curve becomes more linear). \SV{}{We can also observe that 
when $\M$ is less than $0.20$, the relative errors of MP2 are always below $25 \%$ (as denoted by dashed lines in the Figure~\ref{fig_S}).} 
With even a slightly higher $\M$ (around $0.25$), MP2 errors are skyrocketing (up to $80 \%$) and here we encounter stacking complexes, for which MP2 failures are well-known. On the other hand, for hydrogen bonded systems $\M$ values approach $0$, the AC interaction curves become more linear and consequently the MP2 becomes more accurate. In Figure~\ref{fig_C6}, we show the benzene dimer and the acetic acid dimers AC curves obtained by eq~\ref{eq:SPL}, representing a situation when MP2 is accurate (the latter case) and when it is not (the former case). 
 \begin{figure}
    \includegraphics[width=0.95\linewidth]{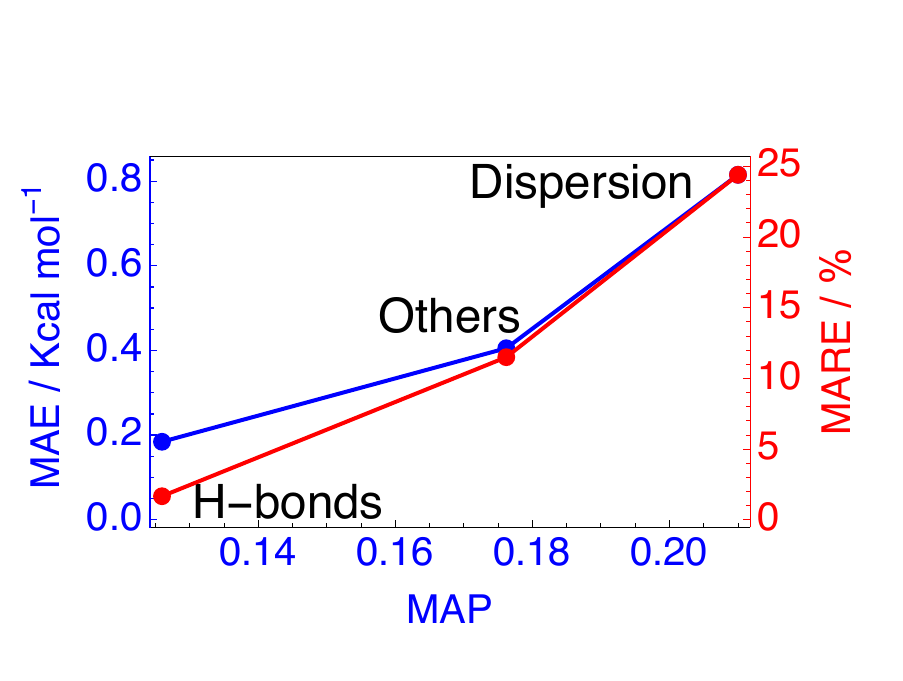}\par
\caption{MARE and MAE for the MP2 method as a function of averaged $\X^{\rm SPL}$
for the subsets of the S66 dataset}
\label{fig_M}
\end{figure}
We also calculate MAE and MARE for the three subsets of the S66 dataset, and these are shown in Figure~\ref{fig_M}, as a function of averaged $\M$ pertaining to a given subeset. As expected, $\M$ increases as we go
from H-bonded complexes, over complexes classified as ``others'' (those bonded by a combination of dispersion and electrostatics)
 to complexes bonded by dispersion. This indicates that the accuracy of MP2 also increases in this order. 

\section{Conclusions and outlook}

In summary, we 
use the AC insights to better understand and predict the accuracy of PT2 theories. 
We also report the highly accurate HF AC curves for the helium isoelectronic series and compare them with their DFT counterparts. While the exact DFT AC curves have been studied extensively in the literature\cite{ColSav-JCP-99,TeaCorHel-JCP-09,TeaCorHel-JCP-10,locpaper}, to the best of our knowledge the HF AC curves (Figure~\ref{fig_hhe}) are reported for the first time here.  

We transform the exact form of our $\X$ indicator into a practical tool  ($\M$) for predicting the accuracy of the MP2 method for systems that dissociate into fragments with non-degenerate ground states. An important point to note about the $\M$ predictor is that it practically comes at no additional computational cost. Computing it by means of eqs~\ref{eq:x1spl} and \ref{eq:1spl}  requires only (beyond the MP2 calculation itself) $W_\infty^{\rm PC}[\rho^{\rm HF}]$, which is easily computed from $\rho^{\rm HF}(\rv)$ and its gradient.  \SV{}{This practical aspect of the $\M$ predictor, combined with its relevance for noncovalent interactions (NIs) and the popularity of MP2 methods for NIs, is even more useful in the light of recent findings of Furche and co-workers.\cite{nguyen2019divergence} Namely, these authors have found that the performance of MP2 for NIs systematically worsens with the increase of a molecular size. Thus they advise caution when MP2 is used for calculating NIs between large molecules, given that the results  can be even qualitatively wrong. This is where our $\M$ predictor can come into play, as it can gauge the reliability of such calculations (as shown in Figure~\ref{fig_S}.)} 
 
The $\M$ indicator is presently applicable to systems that dissociate into fragment with non-degenerate ground states. To address this , we will obtain exact AC curves for small covalently bonded diatomics, and then we will use these curves to get hints on how to transform the exact $\X$ into a practical indicator that also works for systems that dissociate into fragments with degenerate ground states. 

In future work we will also explore the possibility of defining and analyzing the local HF AC curves, as it has been done for their DFT counterparts.\cite{IroTea-MP-15,locpaper,VucIroWagTeaGor-PCCP-17} This could prove useful in using the $\X$ indicator locally (i.e. at a given point in space).

\section{Acknowledgments}\label{ack}
SV acknowledges funding from the Rubicon project (019.181EN.026), which is financed by the Netherlands Organisation for Scientific Research (NWO). KB acknowledges funding from NSF (CHE 1856165). PGG acknowledges funding from the European Research Council under H2020/ERC Consolidator Grant corr-DFT [Grant Number 648932] and from NWO under Vici grant 724.017.001.

\appendix
\section{The PC model and $W_\lambda^{\rm HF}$}
\label{sec_pc}
\begin{figure}
\includegraphics[width=1\linewidth]{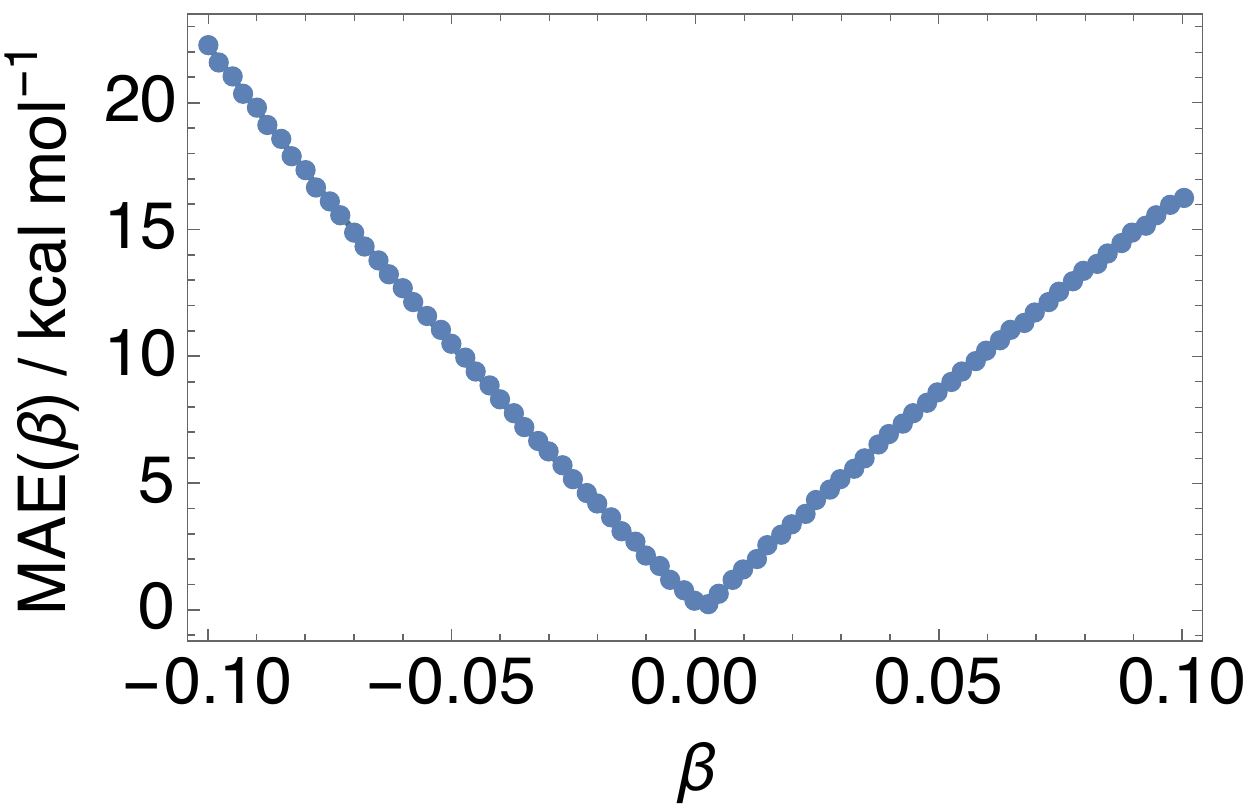}\par
\caption{MAE for the S66 dataset of the SPL interepolation scheme as a function of $\beta$, where we set: $W_\infty^{\rm HF}[\rho^{\rm HF}] =W_\infty^{\beta}[\rho^{\rm HF}]$ (eq~\ref{eq:beta})}
\label{fig_beta}
\end{figure}
\SVc{}{As explained in sec~\ref{sec:predictor}, while $W_{c,\lambda}^{\rm SPL,int}(M)$ (l.h.s. of eq~\ref{eq:wcspl} 
is an accurate approximation to $W_{\lambda,c}^{\rm int}(M)$  (r.h.s of eq~\ref{eq:e}) for NIs,\cite{VucGorDelFab-JPCL-18} we do not expect the SPL scheme to accurately approximate the two terms on the r.h.s. of eq~\ref{eq:e}. Comparing the size of the MP2 and CCSD(T) total energies, we expect these two terms to have a concave adiabatic connection curve. On the other hand, we expect $W_{\lambda,c}^{\rm int}(M)$ to be convex (given that MP2 overbinds a vast majority of S22 and S66 complexes).  The SPL AC curve of eq~\ref{eq:SPL} is always convex, and thus if the two terms on r.h.s. of eq~\ref{eq:e} are concave that would be missed by the  SPL interpolation.  Thus, the accuracy of the  $W_{c,\lambda}^{\rm SPL,int}(M)$ curve for NIs,\cite{VucGorDelFab-JPCL-18} results from an error cancellation between the complex and the monomers. A similar error cancellation has been observed for the fixed-node error in Quantum Monte Carlo calculations of NIs.\cite{DubMitJur-CR-16}}

In this same light we discuss in more details the use of $ W_\infty^{\rm PC} [\rho^{\rm HF}]$ in the SPL interpolation scheme as an approximation to $W_\infty^{\rm HF}[\rho^{\rm HF}]$. 
First we note that the exact $W_\infty^{\rm HF}[\rho^{\rm HF}]$ is expected to be much lower than $ W_\infty^{\rm PC} [\rho^{\rm HF}]$. This is beacause $W_\infty^{\rm PC} [\rho]$ is energetically very close to the exact $W_\infty^{\rm DFT} [\rho]$ (see refs~\onlinecite{SeiGorSav-PRA-07,VucWagMirGor-JCTC-15}), while the following inequality holds\cite{SeiGiaVucFabGor-JCP-18}
\beq \label{eq:ineq}
W_\infty^{\rm HF}[\rho^{\rm HF}] \leq W_\infty^{\rm DFT} [\rho^{\rm HF}] + 2 E_x^{\rm DFT} [\rho^{\rm HF}]
\eeq 
Thus one can even think of approximating $W_\infty^{\rm HF}$ with 
$W_\infty^{\beta =2 }$, where 
\beq \label{eq:beta}
W_\infty^{\beta} [\rho^{\rm HF}] =   W_\infty^{\rm PC} [\rho^{\rm HF}] + \beta E_x^{\rm DFT} [\rho^{\rm HF}].
\eeq
Despite this reasoning, we show here that the use of the ``bare'' PC model (i.e. $W_\infty^{\beta=0 }[\rho^{\rm HF}]$) when used in the SPL interpolation scheme gives more accurate interaction energies. We illustrate this in Figure~\ref{fig_beta}, which shows the MAE for the S66 dataset of the SPL interpolation varies with $\beta$, when we set $W_\infty^{\rm HF}[\rho^{\rm HF}]=W_\infty^{\beta}[\rho^{\rm HF}]$. The minimum in Figure~\ref{fig_beta} \SV{}{lies very close to $\beta=0$. Precisely, it is at $\beta= 0.0016$; (MAE =0.27 kcal/mol) and at $\beta=0$  the MAE is  0.35 kcal/mol.  The error rapidly increases as we go away from this minimum in either of the directions, and already at $|\beta| > 0.01$ the error becomes huge.} By exploring the exact HF AC curves in future work, we will try to better understand why the PC model works so well here  \SV{}{(i.e. why the accuracy of the SPL interpolation is lost if  a $W_\infty[\rho^{\rm HF}]$ lower or higher than $W_\infty^{\rm PC} [\rho^{\rm HF}]$ is used).} We can only make some speculative remarks at this stage. The first one is that we are constructing {\it de facto} an interpolation for interaction energies, and so in some way the PC model is capturing what is needed in this context. After all, we are using the difference between two convex curves to approximate a (probably) convex curve resulting from the difference between two curves that change convexity. Also, we are not using here
the next leading term in the large-$\lambda$ expansion of $W_{c,\lambda}^{\rm HF}$, appearing at orders $\lambda^{-1/2}$,\cite{SeiGiaVucFabGor-JCP-18} which is positive (as in DFT),\cite{GorVigSei-JCTC-09} and expected to be much larger than the DFT one, because, besides zero-point oscillations, it also contains the effect of the operator $-\hat{K}$.\cite{SeiGiaVucFabGor-JCP-18}  When using only  the leading term at large $\lambda$, $W_\infty^{\rm HF}$ must effectively take into account also the positive (large) zero-point term.  For this reason it is not so surprising that a smaller (in absolute value)  $W_\infty^{\rm HF}$ gives better results than the full one. However, it reamins a very intriguing fact the accuracy of the PC model in this context.

\bibliography{all}

\end{document}